%% file: main.tex
\documentclass[sigconf, 9pt]{acmart}

\usepackage{algorithmic}
\usepackage{graphicx}
\usepackage{textcomp}
\usepackage{xcolor}

\usepackage{url}
\usepackage{hyperref}
\usepackage{soul}

\usepackage{array}

\usepackage{float}
\floatstyle{plaintop}
\restylefloat{table}
\usepackage{xspace}
\usepackage{caption}
\usepackage{subcaption}

\usepackage{enumitem}

\usepackage[acronyms,nonumberlist,nopostdot,nomain,nogroupskip,acronymlists={hidden}]{glossaries}
\newglossary[algh]{hidden}{acrh}{acnh}{Hidden Acronyms}
\glsdisablehyper
\input{acronyms.tex}

\setcopyright{acmcopyright}
\copyrightyear{2024}
\acmYear{2024}
\acmDOI{XXXXXXX.XXXXXXX}

\newcommand{\ran}{\gls{ran}\xspace}
\newcommand{\name}{ORANSlice\xspace}

\acmConference[ACM Open AI RAN '2024]{The 1st ACM workshop on Open and AI RAN 2024}{November 18th, 2024}{Washington, D.C., USA}

\usepackage{needspace}


\usepackage{tikzpagenodes,etoolbox}
\usetikzlibrary{calc}
\usepackage[contents={}]{background}
\AddEverypageHook{%
\ifnumequal{\thepage}{1}{%
    \tikz[remember picture,overlay]{%
        \node[draw,
        minimum width=1.03\textwidth,
        text width=1.02\textwidth,
        font=\footnotesize
        ]
        at ($(current page header area) - (0,-20pt)$)
        {%
        This paper has been accepted for publication on the Proceedings of the 1st ACM Workshop on Open and AI RAN (Open-AI RAN '24). This is the author's accepted version of the article. The final version published by ACM is H. Cheng, S. D’Oro, R. Gangula, S. Velumani, D. Villa, L. Bonati, M. Polese, T. Melodia, G. Arrobo, C. Maciocco, ``ORANSlice: An Open-Source 5G Network Slicing Platform for O-RAN'' \textit{Open-AI RAN '24: Proceedings of the 1st ACM Workshop on Open and AI RAN}, Washington, D.C., USA, November 2024.
        };
    }%
}{}
}

\begin{document}
\title{\name: An Open-Source 5G Network Slicing Platform for O-RAN}


\author{Hai Cheng$^{\dagger}$, Salvatore D'Oro$^{\dagger}$, Rajeev Gangula$^{\dagger}$, Sakthivel Velumani$^{\dagger}$,  Davide Villa$^{\dagger}$,  Leonardo Bonati$^{\dagger}$, Michele Polese$^{\dagger}$,  Gabriel Arrobo$^*$, Christian Maciocco$^*$, Tommaso Melodia$^{\dagger}$}
\affiliation{%
\institution{$^{\dagger}$ Institute for the Wireless Internet of Things, Northeastern University, Boston, MA, USA}
\country{}
}
\affiliation{%
\institution{$^*$ Intel Labs, Intel Corporation, Santa Clara, CA, USA}
\country{}
}

\renewcommand{\shortauthors}{H. Cheng et al.}

\begin{abstract}
Network slicing allows \glspl{to} to support service provisioning with diverse \glspl{sla}. 
The combination of network slicing and Open \gls{ran} enables \glspl{to} to provide more customized network services and higher
commercial benefits.
However, in the current Open \gls{ran} community, an open-source end-to-end slicing solution for 5G is still missing. 
To bridge this gap, we developed \name, an open-source network slicing-enabled Open \gls{ran} system integrated with popular open-source \gls{ran} frameworks.
\name features programmable, 3GPP-compliant \gls{ran} slicing and scheduling functionalities. It supports \gls{ran} slicing control and optimization via xApps on the near-real-time \gls{ric} thanks to an extension of the E2 interface between \gls{ric} and \gls{ran}, and service models for slicing.
We deploy and test \name on different O-RAN testbeds and demonstrate its capabilities on different use cases, including slice prioritization and minimum radio resource guarantee. 
\end{abstract}

\maketitle

\glsresetall

\section{Introduction}

Network slicing has been identified as a key technology to deliver bespoke services and superior performance in \glspl{5g}. 
Specifically, \gls{ran} slicing makes it possible to dynamically allocate a certain amount of \gls{ran} resources, e.g., \glspl{prb}, to each slice based on their \gls{qos}, current network conditions, and traffic load. 

The importance of network slicing is emphasized by the O-RAN ALLIANCE, which has defined it as a critical use case and technology in the context of Open \gls{ran} systems~\cite{oran-wg1-slicing-architecture}. The O-RAN architecture foresees slicing in the \gls{ran} and controlled through an xApp deployed on the near-real-time (Near-RT) \gls{ric}. The latter is connected to the \gls{ran} through the E2 interface, whose functionalities can be specified through \emph{\glspl{sm}}. For slicing, the E2SM Cell Configuration and Control (E2SM-CCC)~\cite{oran-wg3-ccc} allows for near-real-time adaptation of slicing parameters via xApps~\cite{oran-wg3-ricarch}. 

Despite the potential of network slicing, which has generated tremendous momentum and technological advancements, practical deployment of \gls{ran} slicing in commercial \gls{5g} networks remains largely unrealized. Numerous studies and research have demonstrated the benefits of \gls{ran} slicing, exploring various aspects, including the use of optimization~\cite{doro2020tnet} and \gls{ai}-based solutions~\cite{zhang2022federated}. 
However,
most of the existing implementations of RAN slicing are confined to research environments and bench setups. 

\gls{ran} software stacks, such as 
\gls{oai}~\cite{oai5g} and srsRAN~\cite{srsRAN}, have been instrumental in advancing the field by welcoming contributions from the open-source community, for example, to integrate and release network slicing technologies for both 5G~\cite{FlexSlice2023} and 4G~\cite{leo2021scope} systems.
Examples of this are provided by SCOPE~\cite{leo2021scope} and FlexSlice~\cite{FlexSlice2023}. The former is a network slicing framework based on srsRAN 4G, which, however, uses custom \glspl{sm} to communicate with the xApps.
%
The latter is a \ran slicing framework with a recursive radio resource scheduler based on \gls{oai} 5G. However, only a single slice can be associated to each \gls{ue}, and it lacks integration with the O-RAN E2SM.


\noindent \textbf{Contribution and Novelty.}~We advance the state-of-the-art by developing and implementing network slicing models compliant with the \gls{3gpp} and O-RAN specifications.
\textit{First}, we extend the 5G protocol stacks of \gls{oai} to support \ran slicing. We extend the original proportional-fair scheduler for \gls{ue} scheduling to a two-tier radio resource scheduler for \ran slices and \glspl{ue}. 
We also develop multi-\gls{pdu} support for the \gls{oai} softwarized 5G \gls{ue}, which enables multiple concurrent slices on the same \gls{ue}.
\textit{Second}, we implement an E2SM-CCC-based \gls{sm} and xApp for \ran slicing control.
The implemented \gls{sm} is O-RAN-compliant and is aligned with 3GPP specifications to enable closed-loop control for \ran slicing via xApps in the Near-RT \gls{ric}.
\textit{Third}, to ensure the robustness and effectiveness of our implementation,
we conduct extensive testing and validation on Arena \cite{bertizzolo20arena}, a SDR-based wireless testbed located in an office environment, and X5G \cite{villa2024x5g}, an O-RAN-compliant production-ready private 5G network testbed.
We show that our implementation can be used to enforce and control slicing policies with both \gls{cots} 5G modules and softwarized \glspl{ue}, as well as different radio devices.
We also evaluate our implementation on different O-RAN \glspl{ric} including the Aether SD-RAN and 
the \gls{osc} Near-RT \glspl{ric}.

By releasing \name to the community,
we hope to bridge the gap between research and practice, providing a valuable resource for further exploration and development.
\footnote{\href{https://github.com/wineslab/ORANSlice}{https://github.com/wineslab/ORANSlice}}

\begin{figure}[t!]
    \centering
    \includegraphics[width=1\linewidth]{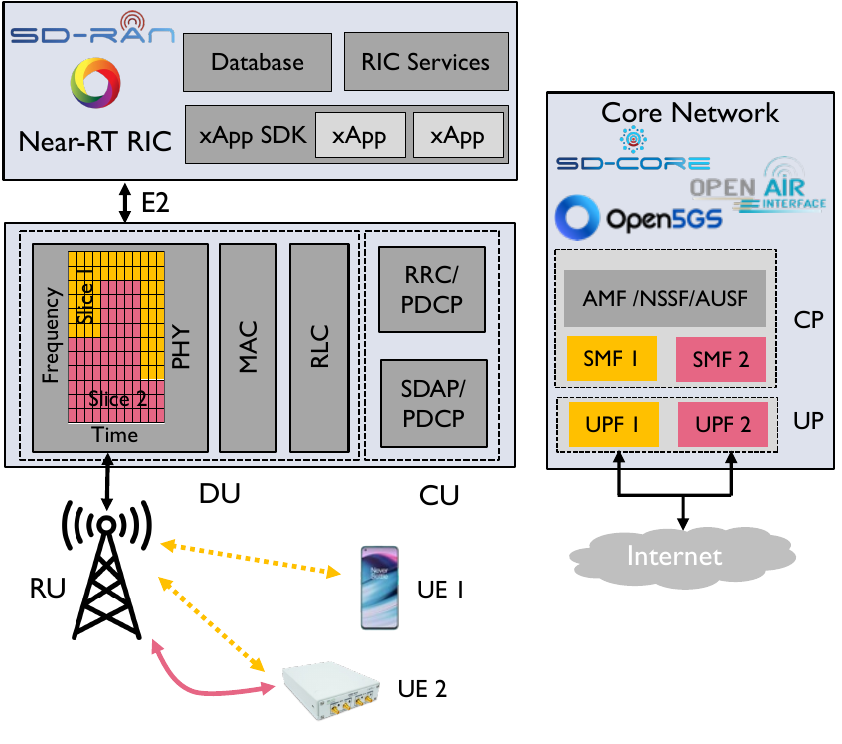}
    \caption{End-to-end network slicing in O-RAN.}
    \label{fig:e2e_netslicing}
\end{figure}

\noindent \textbf{Related Work.} 
 Efforts similar to ours have been documented in the literature. 
ProSlice in \cite{ProSlice2022} developed a customized E2SM and xApp to support \ran slicing. However, ProSlice does not follow the \gls{3gpp} \ran slicing model and the E2SM is not O-RAN-compliant. Besides, ProSlice is not open-source.
RadioSaber \cite{RadioSaber2023nsdi} is a channel-aware \ran slicing framework implemented in a \ran simulator.
Finally, Zipper \cite{Zipper2024nsdi} is a \ran slicing framework based on closed-source commercial protocol stacks.

The closest works to ours are \cite{ProSlice2022} and \cite{FlexSlice2023}. 
Compared to these works, \name advances the state-of-the-art by introducing support for multi-\gls{pdu} and thus multi-slice management on the same \gls{ue}.
This enhancement allows for more efficient and flexible handling of different types of traffic, catering to the diverse needs of modern applications where diverse classes of traffic with different \gls{qos} requirements coexist on the same device and need to be satisfied simultaneously.
\name will be publicly released upon acceptance of this paper, providing the research community with a \gls{3gpp}- and O-RAN-compliant implementation of \gls{ran} slicing on a 5G protocol stack.
In this way, \name can be used to design novel 5G and network slicing functionalities, and test them with \glspl{sdr}, commercial \glspl{ru}, or through the RFSim simulator, which allows \gls{oai} \gls{gnb} and \gls{ue} to communicate over an Ethernet connection.%


\vspace{-0.1cm}
\section{O-RAN and Network Slicing}
\label{sec:openran_framework}

Before providing details on the implementation of \name, in the following we provide useful background on Open \ran, network slicing (with a specific focus on \ran slicing, as defined by the 3GPP~\cite{3gpp.28.541}), and its role in O-RAN. 
The high-level system architecture of our end-to-end network-slicing solution is illustrated in Figure~ \ref{fig:e2e_netslicing}. First, we introduce the network architecture, with \gls{ran}, Near-RT \gls{ric}, and \gls{cn}. Then, we elaborate on how network slicing is supported by the three sub-systems. Finally, we present \ran slicing within the O-RAN architecture.

\vspace{-0.1cm}
\subsection{O-RAN Architecture}

As shown in Figure~\ref{fig:e2e_netslicing}, Open \gls{ran} \glspl{gnb} are disaggregated into an \gls{ru}, a \gls{du}, and \gls{cu}~\cite{polese2023understanding}. These elements connect to a core network with multiple \glspl{nf} managing mobility, session, authentication, billing, and routing to the public Internet.
O-RAN also introduces the concept of \glspl{ric}, i.e., software components that host applications to monitor, control and optimize \gls{ran} functions~\cite{oran-wg3-ricarch}. 
The Non-RT \gls{ric} is responsible for the optimization and control of \gls{ran} functions (or resources) on a time scale larger than 1 second via rApps. 
The Near-RT \gls{ric} is responsible for \gls{ran} optimization and control via xApps with a time requirement between 10 milliseconds to 1 second. 
Via xApps and rApps, the \gls{ric} leverages \glspl{kpm} and \gls{ai} algorithms to compute control policies and enforce actions on the \gls{ran}. 


\vspace{-0.1cm}
\subsection{End-to-end Network Slicing} \label{sec:e2e_slicing}

A network slice is an end-to-end logical network spanning both \gls{ran} and \gls{cn}. It can be dynamically created and configured to provide bespoke services to serve a diverse set of applications and use cases.
In a cellular network, each network slice is uniquely identified by a \gls{snssai}, which consists of a \gls{sst} and a \gls{sd}.  
\gls{sst} is an 8-bit mandatory field identifying the slice type. 
\gls{sd} is a 24-bit optional field that differentiates among slices with the same \gls{sst}. 
It is worth mentioning that the same \gls{ue} could be subscribed to up to 8 slices, which benefits applications with different network requirements.


\noindent \textbf{RAN Slicing.}~Radio resources are organized in a time and frequency grid (Figure~\ref{fig:e2e_netslicing}). Each grid element is referred to as a \gls{prb} and is used to schedule \gls{ue} transmissions and broadcast control messages, among others.  
Network slicing in the \gls{ran} (i.e., \gls{ran} slicing), refers to the problem of allocating (and dedicating) such radio resources (i.e., \glspl{prb}) to different slices according to certain slicing policies. 


\noindent \textbf{Core Network Slicing.}
The \gls{5g} \gls{cn} enables secure and reliable connectivity between the \gls{ue} and the Internet. As shown in Figure~ \ref{fig:e2e_netslicing}, the \gls{5g} \gls{cn} is decoupled into \gls{up} and \gls{cp}.
To enable network slicing in the \gls{cn} (i.e., core slicing), dedicated \glspl{nf} in both \gls{cp} and \gls{up} need to be created for each slice. For example, a dedicated \gls{smf} and \gls{upf} pair can be created for each slice (Figure~\ref{fig:e2e_netslicing}). 
The other \glspl{nf} in the \gls{cn}, such as the \gls{amf}, can instead be shared.

\noindent \textbf{Multi-slice Support.}
\gls{3gpp} specifies a set of procedures to link \gls{ue} traffic to a certain network slice. Specifically, during the \gls{pdu} session establishment phase, the \gls{ue} can specify the \gls{snssai} of the target network slice.
Upon establishment, the \gls{pdu} session is assigned an IP, allowing applications in the \gls{ue} to access the network service of the network slice by binding to that IP. 
This procedure makes it possible to establish multiple \gls{pdu} sessions on the same \gls{ue}, and each \gls{pdu} can be bound to a dedicated network slice tailored to the considered use case or application executed by the \gls{ue}.  

\vspace{-0.1cm}
\subsection{RAN Slicing in O-RAN}

The O-RAN WG3 has identified \ran slicing as a key use case in the context of the Near-RT \gls{ric} specifications~\cite{oran-wg3-ucr}. It has also released the O-RAN service model E2SM-CCC~\cite{oran-wg3-ccc} which details the structure and procedures necessary to enable \ran slicing in O-RAN following the 3GPP specifications described above. This is performed via xApps executing at the Near-RT \gls{ric}, where the limited radio resources are managed in near-real-time according to \gls{qos} requirements and highly varying \ran load and conditions.

\vspace{-0.1cm}
\section{\name Implementation}
\label{sec:main_func}

This section details the design and implementation for the missing blocks required to support the \ran slicing use cases based on the O-RAN specifications.

From a software point of view, the open-source ecosystem already offers all architectural blocks necessary to instantiate and operate a 5G network.
For example, disaggregated 5G base stations can be instantiated via \gls{oai} \cite{oai5g} and srsRAN \cite{srsRAN}, which also offer O-RAN integration and functionalities. A 5G core network can be instantiated via \gls{oai}, Open5GS \cite{open5gs}, or SD-Core~\cite{onf}, all of which support core slicing. Similarly, the \gls{osc} and Aether offer open-source implementations of Near-RT \glspl{ric}. OpenRAN Gym, an open-source project for collaborative research in the O-RAN ecosystem, provides components to connect across \gls{ran} and \glspl{ric}~\cite{bonati2022openrangym_pawr}. 


What is missing is an open-source, 3GPP- and O-RAN-compliant implementation of \ran slicing functionalities and the support for multi-slice applications at the same \gls{ue}. 
In this paper, we design and develop \name, an open-source extension to \gls{oai} to fill the gap between O-RAN and end-to-end network slicing. 


\vspace{-0.1cm}
\subsection{RAN Slicing Enabled Protocol Stacks}

The protocol stack of \name is based on \gls{oai}~\cite{oai5g}. 
To enable \gls{ran} slicing, \name advances and extends the functionalities of the \gls{mac} layer, including slice information of \gls{pdu} sessions of each \gls{ue} and a re-designed two-tier radio resource scheduler.  
Moreover, we extend the \gls{oai} nrUE to support the instantiation and management of multiple \gls{pdu} sessions for different slices.


\begin{figure}
    \centering
    \includegraphics[width=1\linewidth]{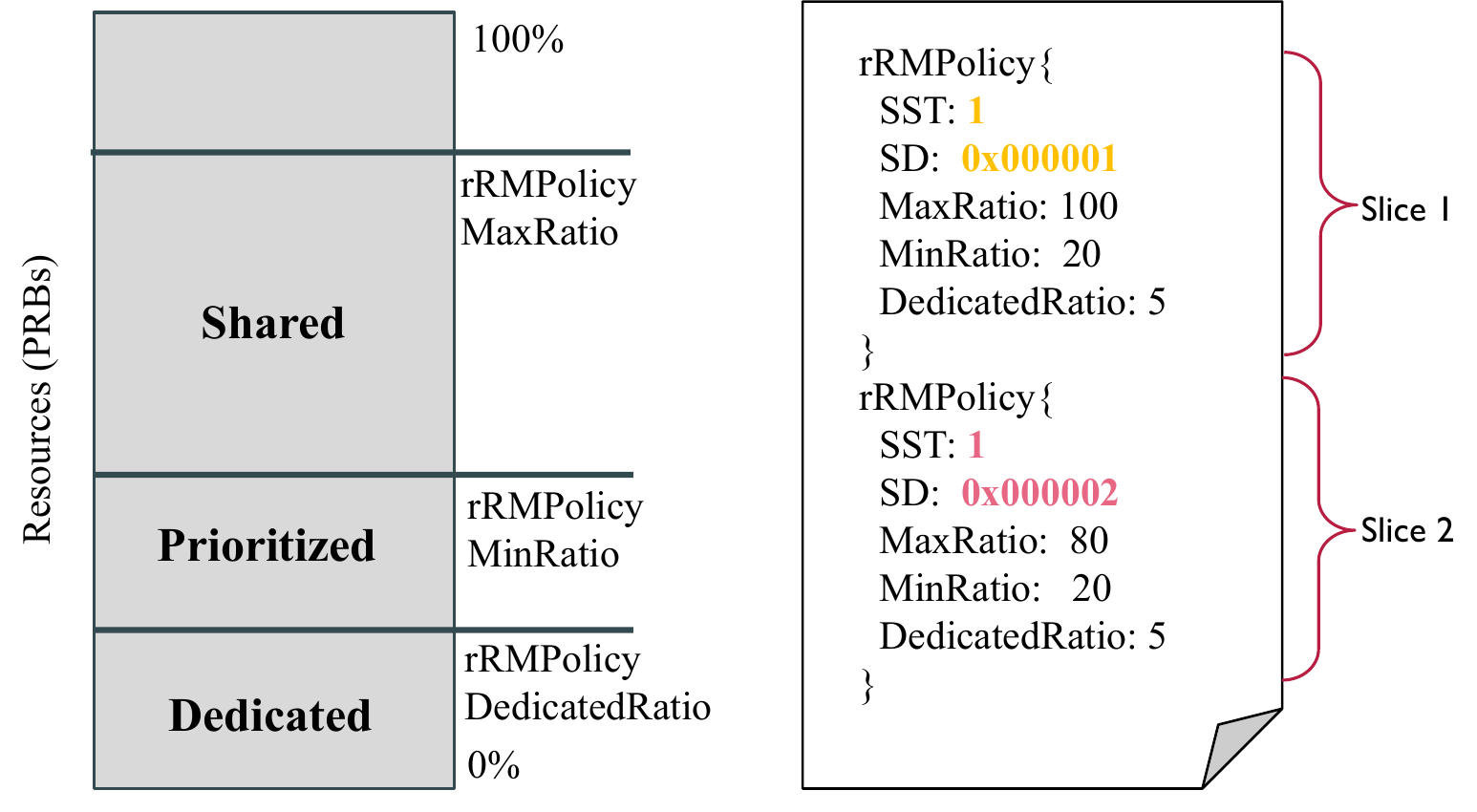}
    \caption{Illustration of RRM Policy Ratio Message}
    \label{fig:rrmPolicy}
\end{figure}

\noindent \textbf{Slice-Aware MAC.} 
In \gls{oai}, the \gls{mac} layer implements a \emph{proportio\-nal-fair} scheduling algorithm to allocate radio resources between different \glspl{ue}.
However, to realize \gls{ran} slicing, the \gls{mac} scheduler needs to be aware of \gls{pdu}-slice associations so as to properly allocate resources among users.

There are two basic types of \ran slicing schemes. The first one implements \textit{slice-isolation}, where a fixed amount of resources are exclusively dedicated to each slice to guarantee resource availability~\cite{leo2021scope}. This method completely prevents resource sharing, has a low resource utilization efficiency, as unused \glspl{prb} are not reallocated to other slices, but has a low-complexity implementation.
The second one is the \textit{slice-aware} scheme, where resources can be exclusively allocated to each slice, or shared among slices with a priority-based access mechanism. The slice-aware scheme increases the resource utilization efficiency as \glspl{prb} allocated to a slice, but unused, can be distributed across other slices on demand.

Since the second approach is more efficient, it has been selected by the 3GPP~\cite{3gpp.28.541} and the O-RAN ALLIANCE~\cite{oran-wg1-slicing-architecture} as a reference implementation for \ran slicing and, for this reason, we consider this in \name. 
In the context of \ran slicing, the 3GPP introduces the concept of \gls{rrm} \gls{ran} slicing policy~\cite{3gpp.28.541}, as illustrated in Figure~\ref{fig:rrmPolicy}. The \texttt{rRMPolicyDedicatedRatio} represents the dedicated percentage of \glspl{prb} allocated to the network slice and is exclusively allocated to the slice even if the slice has no active traffic demand. 
The 3GPP also defines the concept of prioritized \gls{prb} access via two parameters: \texttt{rRMPolicyMinRatio} and \texttt{rRMPolicy\-Max\-Ratio}.
The former represents the minimum percentage of \glspl{prb} that would be prioritized for allocation to the network slice.
The latter represents the maximum percentage of \glspl{prb} that can be allocated to the slice. At a high level, the percentage of \glspl{prb} that falls within the prioritized range (i.e., defined as the difference between the maximum ratio and the dedicated ratio) is guaranteed to be allocated to the slice, but only if the slice has active users requesting \glspl{prb}. 

\name extends the \emph{proportional-fair} scheduler of \gls{oai} to integrate the slice-aware scheme described above. As we will describe later, this is achieved by implementing a two-tier resource allocation mechanism that considers: (i) inter-slice resource allocation and sharing according to \gls{ran} slicing policies, which will be elaborated in Section~\ref{sec:e2sm_rrmpolicy}; and (ii) intra-slice resource allocation to schedule transmissions for all \glspl{ue} that belong to the same slices.\footnote{We would like to mention that support for the dedicated ratio defined by the \gls{rrm} policy in Figure~\ref{fig:rrmPolicy} will be added in the future.}

\noindent \textbf{UE Slicing Support in OAI nrUE.} 
The other important contribution of \name is the introduction of support for multiple slices on the same \gls{ue}. 
Prior to \name, \gls{oai} softwarized \gls{5g} \gls{ue}, i.e., \gls{oai} nrUE, supported only one \gls{pdu} session per \gls{ue}, limiting each \gls{ue} to a single active slice at any time. To enable multi-slice support, we have extended nrUE functionalities to: (i) enable the coexistence of multiple \gls{pdu} sessions on each \gls{ue}; (ii) instantiate/delete new \glspl{pdu} on-demand; and (iii) assign each \gls{pdu} to a slice. 
We already contributed these functionalities to \gls{oai} nrUE repository and they are available to the community as open-source components. It is worth mentioning that currently available 5G smartphones do not support these features, and we hope that this will 
help the community to investigate and explore \gls{ran} slicing topics that go beyond single \gls{ue}-slice associations.

\vspace{-0.1cm}
\subsection{Implementing the E2SM-CCC}
\label{sec:e2sm_rrmpolicy}

To enable the O-RAN \ran slicing use case, we also developed the corresponding E2 \gls{sm} to realize the O-RAN E2SM-CCC service model. We developed two versions of the E2SM-CCC. One has been integrated with the Linux Foundation's Aether SD-RAN Near-RT \gls{ric}~\cite{onf} and already made public,\footnote{\href{https://github.com/onosproject/onos-e2-sm/pull/392}{https://github.com/onosproject/onos-e2-sm/pull/392}}
while a simplified version has been integrated with the \gls{osc} Near-RT \gls{ric}. Due to space limitations, in this paper we describe and present results based on the latter only.

\noindent\textbf{E2SM-CCC Service Model.}
The E2 \gls{sm} describes how a specific \gls{ran} function (the \textit{service}) within an E2 node interacts with the Near-RT \gls{ric} and its xApps. 
The E2 \gls{sm} consists of an \gls{e2ap} and a data schema accepted by both the \gls{ran} function and Near-RT \gls{ric}.
The flexible and programmable E2SM-CCC is critical for the autonomous and intelligent \ran control loop in O-RAN.

To support \ran slicing reconfiguration via xApps, we have implemented a simplified version of the E2SM-CCC service model integrated with the \gls{osc} Near-RT \gls{ric} to support 3GPP-compliant \ran slicing reconfiguration via xApps. 

Only the \textit{O-RRMPolicyRatio} configuration of E2SM-CCC is supported in the implemented service model.
Other configurations defined in E2SM-CCC are omitted as they are beyond the scope of this paper.
Protobuf is used to send control messages that include the \gls{rrm} policy necessary to update \ran slicing strategies enforced by the \ran. We are currently extending our implementation to use ASN.1 definitions to encode/decode the message payload, as well as to adopt the full O-RAN E2SM-CCC specifications.

\vspace{-0.1cm}
\subsection{RAN Slicing xApp and Data-driven Control Automation}

We developed a \gls{ran} Slicing xApp to compute \gls{rrm} policies necessary to update \ran slicing strategies. 
Specifically, the \ran Slicing xApp periodically reads the \gls{ran} \glspl{kpi} from an InfluxDB database, processes them, and generates the slicing policies that are sent to the gNB through the E2 interface via a \gls{ric} Control message. 
Control messages from the xApp are serialized into Protobuf objects and sent via the \glspl{sm} described above. 

The data from the gNB is periodically sent to a dedicated \gls{osc} \gls{kpm} xApp that reads \glspl{kpm} received over the E2 termination through a \gls{ric} Indication message. The \gls{kpm} xApp
inserts the received \gls{ran} \glspl{kpm} into the InfluxDB database that is leveraged by the \ran Slicing xApp. Similarly to the \gls{ric} Control messages, the payload of the \gls{ric} Indication messages is serialized into Protobuf objects.

\vspace{-0.1cm}
\section{Testbeds and Experiment Results}
\label{sec:use_cases}

In this section, we demonstrate \name's portability by deploying an end-to-end O-RAN cellular network on two testbeds: Arena and X5G, discussed in Section~\ref{sec:testbed}.
We demonstrate the functionality of \name by performing two experiments.
In the first experiment (Section~\ref{sec:ran_slicing_ctrl}), we show an end-to-end network slicing deployment with \gls{kpm} and \ran slicing xApp running in the \gls{osc} Near-RT \gls{ric}. The goal of this first experiment is to demonstrate the \ran slicing functionalities enabled by \name.
In the second experiment (Section~\ref{sec:qos}), we show a simple, yet illustrative, example of how to guarantee access to a minimum amount of radio resources via tailored \gls{ran} slicing policies in \name. In this latter experiment, we also demonstrate the multi-slice support offered by \name. 
In addition to the \gls{ota} transmission experiments on two testbeds, we also replicate the two experiments in \gls{oai} RFSim mode to show the ability of \name working without physical \gls{ue} or radio devices.

\vspace{-0.1cm}
\subsection{Multi-vendor O-RAN Testbeds}
\label{sec:testbed}


To demonstrate \name on the aforementioned testbeds, we deploy an end-to-end O-RAN 5G network system that matches the architecture illustrated in Figure~\ref{fig:e2e_netslicing}. This system consists of the following four elements: \glspl{ue}, \gls{gnb}, Near-RT \gls{ric}, and \gls{5g} \gls{cn}. 
While we use exactly the same \glspl{ue}, Near-RT \gls{ric} and \gls{5g} \gls{cn} across the \gls{ota} testbeds, Arena and X5G testbed use different gNB architectures as described below. 

\noindent \textbf{Arena Testbed and \gls{gnb}.}
The Arena testbed is a remotely accessible wireless testing platform located in a large indoor office environment.
It features 24 \gls{usrp} X310 and X410 connected to Dell EMC PowerEdge R340 servers via a 10/100 Gbps switch. 
The antennas are hung off the ceiling of an office space and connected to the \glspl{usrp} for a total of 64 antennas in an $8 \times 8$ grid.
For the \gls{5g} \gls{ran} deployed on the testbed, the \gls{usrp} serves as \gls{ru} and the \gls{oai} \gls{5g} protocol stack running on the Dell server serves as \gls{du}/\gls{cu}. We consider a configuration with a bandwidth of $40$~MHz ($106$ \glspl{prb} in $30$~KHz subcarrier spacing). 

\noindent \textbf{X5G Testbed and \gls{gnb}.}
The X5G testbed is a private, multi-vendor \gls{5g} network. 
The \gls{ran} protocol stack consists of: (i) NVIDIA Aerial, a GPU-accelerated L1 \gls{phy} layer; (ii) higher layers implemented via \gls{oai} and running on X86 CPU; and (iii) a Foxconn \gls{ru} operating in the n78 band and connected to the DU via the O-RAN fronthaul. In this testbed, we consider a bandwidth of $100$~MHz with $273$ \glspl{prb} and $30$~KHz subcarrier spacing.

\noindent \textbf{5G UE.}
We test both \gls{cots} and softwarized \glspl{ue}. 
The \gls{cots} Sierra Wireless EM9191 \gls{5g} module is selected to perform flexible experiments since it supports multiple \gls{pdu} sessions and customized \gls{snssai} for each \gls{pdu} session. 
We also tested \name with the \gls{oai} nrUE, a softwarized \gls{ue} that we have extended to enable multi-\gls{pdu} sessions support and thus \gls{ue} multi-slicing.

\noindent \textbf{5G Core Network.}
To validate slicing operations, we tested \name with \gls{oai} \gls{cn}, Open5GS, and SD-Core. These are modular core networks and support network slicing natively. 

\noindent \textbf{Near-RT \gls{ric}.}
The Near-RT \gls{ric} and the xApps of Figure~\ref{fig:e2e_netslicing} are deployed via Red Hat OpenShift. Specifically, we deployed the ``E'' release of the \gls{osc} Near-RT \gls{ric} together with the \gls{osc} \gls{kpm} xApp and the \gls{ran} Slicing xApp. 



\vspace{-0.1cm}
\subsection{Testing RAN Slicing Control}
\label{sec:ran_slicing_ctrl}



In the Near-RT \gls{ric}, the deployed \gls{kpm} xApp and \gls{ran} Slicing xApp work together to enable \gls{ran} slicing control. Every $0.5$ seconds, the \gls{kpm} xApp acquires \glspl{kpi} for all connected \glspl{ue} from the E2 interface. This includes user information (e.g., \gls{rnti} and \gls{snssai}) and \glspl{kpm} such as \gls{bler}, \gls{mcs}, and throughput. This data is stored in an InfluxDB database in the form of time series data. 

The \gls{ran} Slicing xApp reads the \glspl{kpm} from the database and calculates the average downlink throughput for each slice for the previous $5$ seconds. Then, it identifies the slice with the lowest and highest reported throughput and sets their \texttt{rRMPolicyMaxRatio} to $90\%$ and $10\%$, respectively. 
The goal of this experiment is to demonstrate the correctness of the \gls{rrm} policy update. 
How to use \name to satisfy a target \gls{qos} will be shown in Section~\ref{sec:qos}.

\begin{figure}[t!]
    \centering
    \includegraphics[width=1\linewidth]{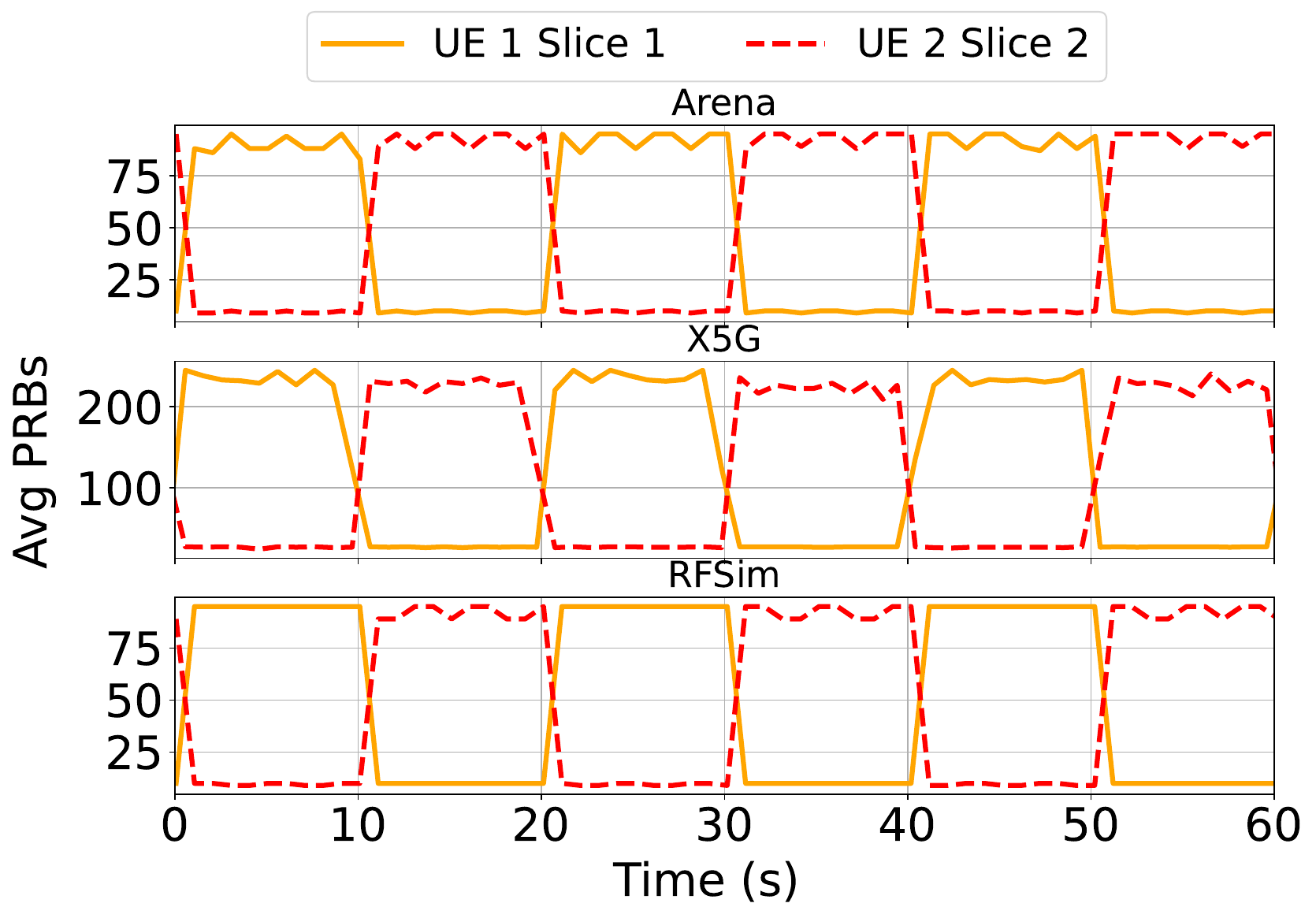}
    \caption{Average number of DL \glspl{prb} allocated to each slice in the first experiment.}
    \label{fig:dl_prb_slicing_ctrl}
\end{figure} 

\begin{figure}[t!]
    \centering
    \includegraphics[width=1\linewidth]{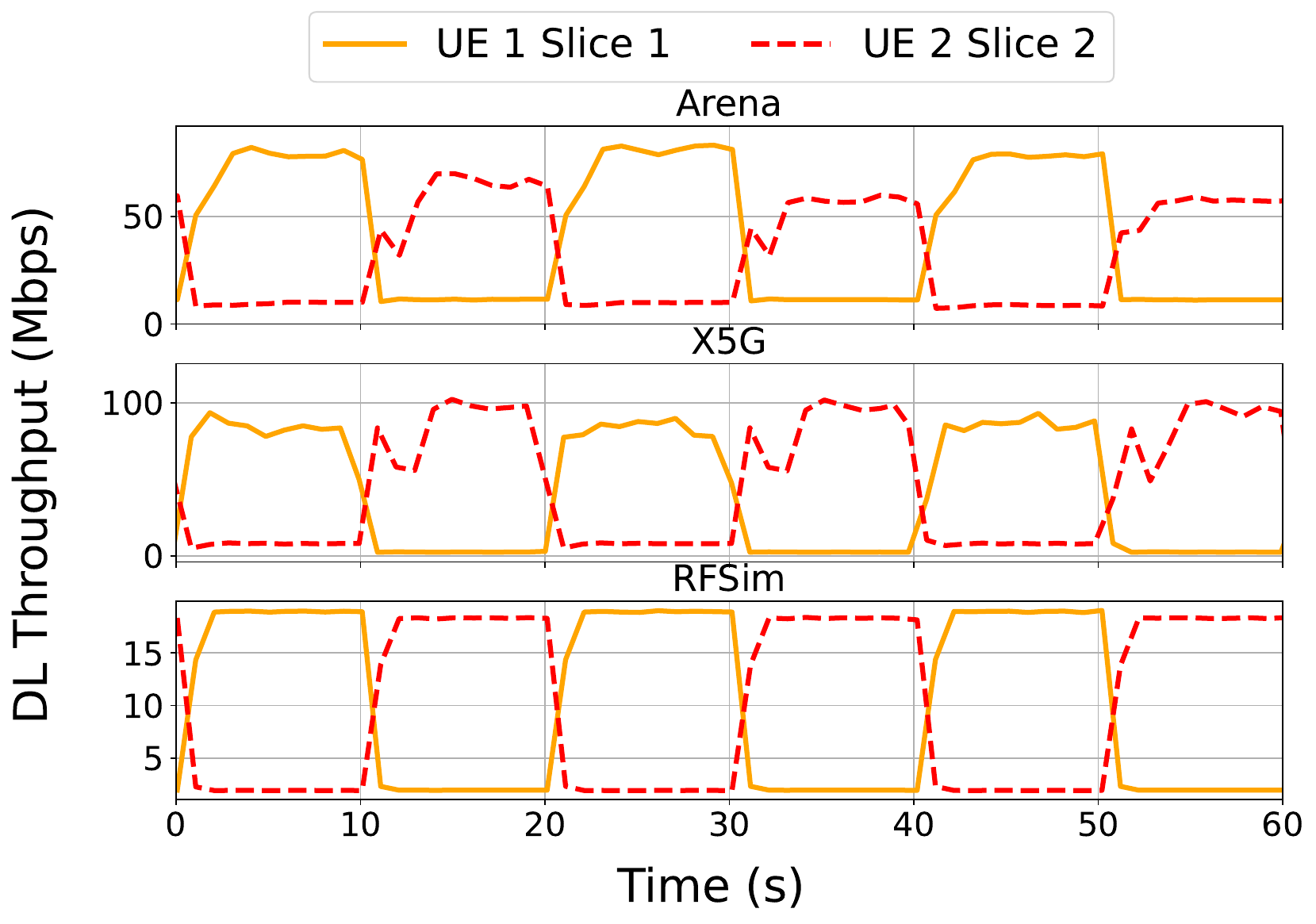}
    \caption{DL Throughput for each slice in the first experiment.}
    \label{fig:dl_thruput_slicing_ctrl}
\end{figure}

We consider two slices and two \glspl{ue}, each associated to one slice. The control logic computes a new \gls{rrm} policy every $10$ seconds. 
As shown in Figure~\ref{fig:dl_prb_slicing_ctrl}, the averaged \glspl{prb} per frame for \gls{ue} 1 and \gls{ue} 2 changes every $10$ seconds. For the Arena testbed (which can allocate up to 106 \glspl{prb}), the \glspl{prb} of each \gls{ue} varies between $10$ and $90$. For the X5G testbed, the \glspl{prb} of each \gls{ue} range from $27$ to $230$ (the gNB in this testbed can allocate up to 273 \glspl{prb}). RFSim has the same 5G numerology as Arena, and the average number of allocated \glspl{prb} follows that of the same pattern.
In all cases, the number of \glspl{prb} allocated to each slice is approximately equal to $10\%$ and $90\%$ of the number of available \glspl{prb} in each testbed, which proves that the \gls{rrm} policy is applied correctly.
In Figure~\ref{fig:dl_thruput_slicing_ctrl}, we also show the impact that the slicing policy has on downlink throughput of each slice, which depends on the amount of \glspl{prb} available to the slice.
We see the downlink throughput for the two \glspl{ue} varies following the number of \glspl{prb} in Figure~\ref{fig:dl_prb_slicing_ctrl}. The difference between the throughput values of \glspl{ue} is related to the differences in the hardware and RF channel peculiar to each testbed.


\vspace{-0.1cm}
\subsection{Multi-Slice with Guaranteed \glspl{prb}} \label{sec:qos}

In this experiment, we show how \name handles multi-slice applications and guarantees minimum \gls{prb} allocation by fine-tuning \texttt{rRMPolicyMinRatio} in the \gls{rrm} policy. 
We consider two slices and 2 \glspl{ue}, and \gls{ue} 1 activates two \glspl{pdu} each with a different \gls{snssai}. This means that UE 1 has active \glspl{pdu} on both slices.  
We set $\texttt{rRMPolicyMinRatio} = 0$ and $\texttt{rRMPolicyMaxRatio} = 100$ for all slices, i.e., no minimum \gls{prb} guarantee.

\begin{figure}[t!]
    \centering
    \includegraphics[width=1\linewidth]{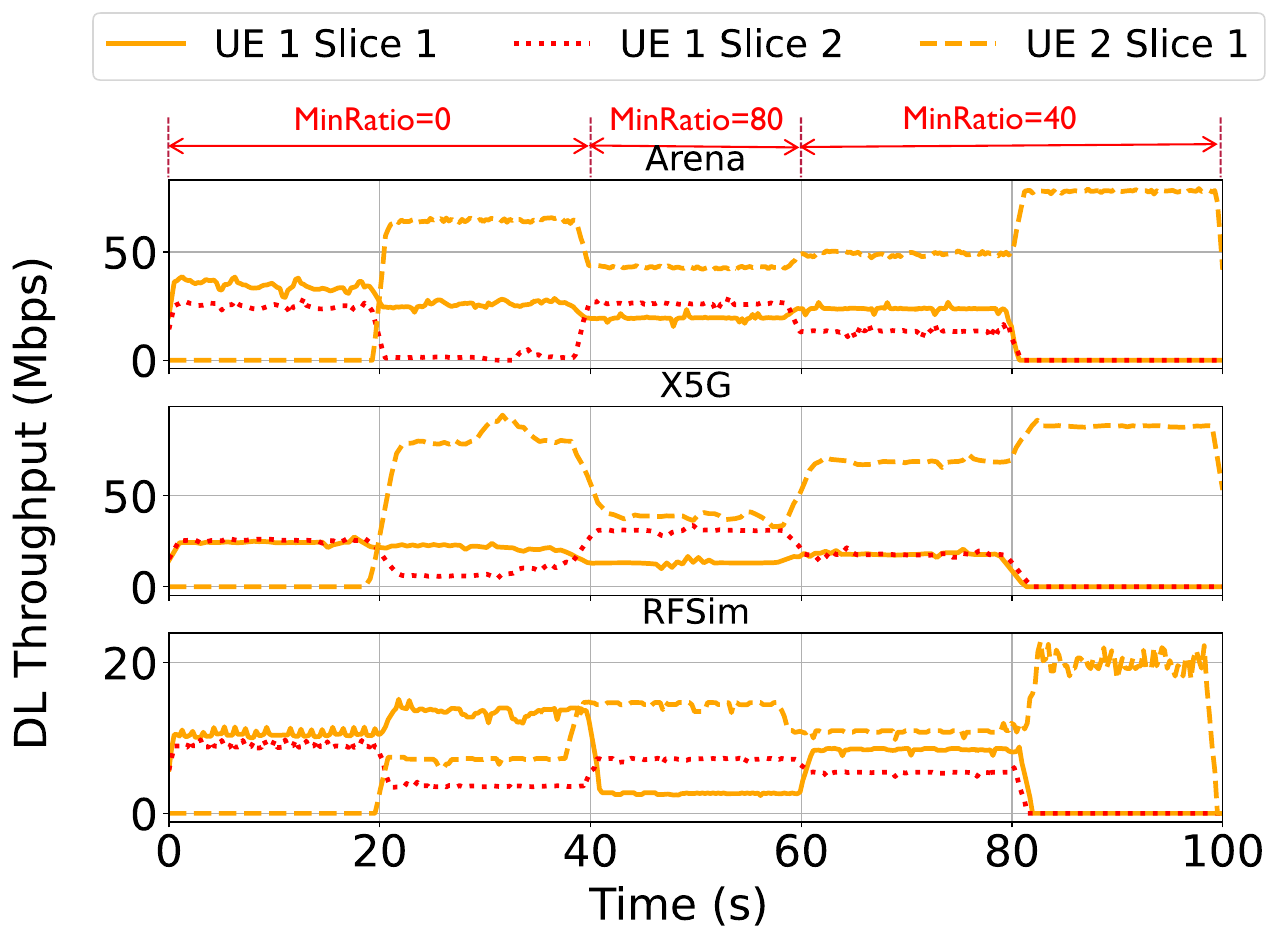}
    \caption{DL throughput for each slice and \gls{ue} in the second experiment.}
    \label{fig:slices_qos}
\end{figure}

The experiment evolution is illustrated in Figure~\ref{fig:slices_qos}. 
In step~1 ($0\sim20$s), \gls{ue} 1 establishes two \gls{pdu} sessions, with one \gls{pdu} session associated with each slice. For each \gls{pdu}, \gls{ue}~1 generates TCP downlink data via \texttt{iPerf3}, while
\gls{ue} 2 is inactive.
We notice that the throughput levels achieved by the two \glspl{pdu} of \gls{ue} 1 are comparable thanks to the proportional-fair scheduler.
In step~2 ($20\sim40$s), \gls{ue} 2 establishes a \gls{pdu} session associated to slice $1$, and starts an \texttt{iPerf3} TCP downlink data transmission. The throughput of \gls{ue} 1's \gls{pdu} associated with slice $2$ decreases to about 1~Mbps, due to the resource competition caused by the new establishment of \gls{ue} 2's PDU on slice $1$. 
Moreover, since $\texttt{rRMPolicyMinRatio} = 0$ for all slices, the gNB does not attempt to improve the throughput of \gls{ue} 1's PDU associated with slice $2$ as the \gls{rrm} policy does not specify any guaranteed \gls{prb} provisioning for slice $2$.
In step~3 ($40\sim60$s), the $\texttt{rRMPolicyMinRatio}$ of slice $2$ is set to $80$, which makes the throughput of \gls{ue} 1's \gls{pdu} associated to slice $2$ increase thanks to the minimum \glspl{prb} guarantee.
Similarly, in step~4 ($60\sim80$s), the $\texttt{rRMPolicyMinRatio}$ of slice $2$ is decreased from $80$ to $40$ and the throughput of \gls{ue} 1 on slice $2$ decreases.
In step~5 (80 to 100 seconds), \gls{ue} 1 stops the \texttt{iPerf3} transmission at the beginning of this step, while \gls{ue} 2 stops it at the end.
This experiment demonstrates how fine-tuning the $\texttt{rRMPolicyMinRatio}$ parameter for each slice can guarantee a minimum \gls{prb} level for individual slices.


\vspace{-0.1cm}
\section{Conclusions}
\label{sec:conclusion}

This paper presents \name, an open-source 5G framework for network slicing in O-RAN. 
\name extends \gls{oai} to deliver \gls{3gpp}-compliant \ran slicing and support for multi-slice applications. It also integrates an E2 service model based on E2SM-CCC, and a \ran Slicing xApp working with the \gls{osc} Near-RT \gls{ric}.
The \ran slicing functionalities of \name are demonstrated on two over-the-air O-RAN testbeds and on the \gls{oai} RF simulator for different use cases. 
We hope that \name can accelerate the \ran slicing research by providing an open-source \gls{3gpp}- and O-RAN-compliant \ran slicing framework.

\section*{Acknowledgment}
This work was supported by the U.S. National Science Foundation
under grant CNS-2117814.

\vspace{-0.1cm}
\bibliographystyle{ACM-Reference-Format}
\bibliography{sample-base}

\end{document}

%% file: acronyms.tex
\newacronym{3gpp}{3GPP}{3rd Generation Partnership Project}
\newacronym{4g}{4G}{4th generation mobile network}
\newacronym[plural=5G,\glsshortpluralkey=5G,firstplural=5th generation mobile networks]{5g}{5G}{5th generation mobile network}
\newacronym[plural=6G,\glsshortpluralkey=6G,firstplural=6th generation mobile networks (6G)]{6g}{6G}{6th generation mobile network}
\newacronym{nextg}{NextG}{Next Generation}
\newacronym{5gc}{5GC}{5G Core}
\newacronym{adc}{ADC}{Analog to Digital Converter}
\newacronym{aerpaw}{AERPAW}{Aerial Experimentation and Research Platform for Advanced Wireless}
\newacronym{ai}{AI}{Artificial Intelligence}
\newacronym{aimd}{AIMD}{Additive Increase Multiplicative Decrease}
\newacronym{am}{AM}{Acknowledged Mode}
\newacronym{amc}{AMC}{Adaptive Modulation and Coding}
\newacronym{amf}{AMF}{Access and Mobility Management Function}
\newacronym{aops}{AOPS}{Adaptive Order Prediction Scheduling}
\newacronym{api}{API}{Application Programming Interface}
\newacronym{apn}{APN}{Access Point Name}
\newacronym{aqm}{AQM}{Active Queue Management}
\newacronym{ausf}{AUSF}{Authentication Server Function}
\newacronym{avc}{AVC}{Advanced Video Coding}
\newacronym{awgn}{AGWN}{Additive White Gaussian Noise}
\newacronym{balia}{BALIA}{Balanced Link Adaptation Algorithm}
\newacronym{bbu}{BBU}{Base Band Unit}
\newacronym{bdp}{BDP}{Bandwidth-Delay Product}
\newacronym{ber}{BER}{Bit Error Rate}
\newacronym{bf}{BF}{Beamforming}
\newacronym{bler}{BLER}{Block Error Rate}
\newacronym{brr}{BRR}{Bayesian Ridge Regressor}
\newacronym{bsr}{BSR}{Buffer Status Report}
\newacronym{bs}{BS}{Base Station}
\newacronym{bpsk}{BPSK}{Binary Phase-shift keying}
\newacronym{bss}{BSS}{Business Support System}
\newacronym{ca}{CA}{Carrier Aggregation}
\newacronym{caas}{CaaS}{Connectivity-as-a-Service}
\newacronym{cb}{CB}{Code Block}
\newacronym{cc}{CC}{Congestion Control}
\newacronym{ccc}{CCC}{Cell Configuration and Control}
\newacronym{ccid}{CCID}{Congestion Control ID}
\newacronym{cco}{CC}{Carrier Component}
\newacronym{cd}{CD}{Continuous Delivery}
\newacronym{cdd}{CDD}{Cyclic Delay Diversity}
\newacronym{cdf}{CDF}{Cumulative Distribution Function}
\newacronym{cdma}{CDMA}{Code-Division Multiple Access}
\newacronym{cdn}{CDN}{Content Distribution Network}
\newacronym{ci}{CI}{Continuous Integration}
\newacronym{cicd}{CI/CD}{Continuous Integration/Continuous Delivery}
\newacronym{cir}{CIR}{Channel Impulse Response}
\newacronym{cn}{CN}{Core Network}
\newacronym{codel}{CoDel}{Controlled Delay Management}
\newacronym{comac}{COMAC}{Converged Multi-Access and Core}
\newacronym{cord}{CORD}{Central Office Re-architected as a Datacenter}
\newacronym{cornet}{CORNET}{COgnitive Radio NETwork}
\newacronym{cosmos}{COSMOS}{Cloud Enhanced Open Software Defined Mobile Wireless Testbed for City-Scale Deployment}
\newacronym{cots}{COTS}{Commercial Off-the-Shelf}
\newacronym{cp}{CP}{Control Plane}
\newacronym{up}{UP}{User Plane}
\newacronym{cpu}{CPU}{Central Processing Unit}
\newacronym{cqi}{CQI}{Channel Quality Information}
\newacronym{cr}{CR}{Cognitive Radio}
\newacronym{cran}{CRAN}{Cloud \gls{ran}}
\newacronym{crs}{CRS}{Cell Reference Signal}
\newacronym{csi}{CSI}{Channel State Information}
\newacronym{csirs}{CSI-RS}{Channel State Information - Reference Signal}
\newacronym{cu}{CU}{Central Unit}
\newacronym{d2tcp}{D$^2$TCP}{Deadline-aware Data center TCP}
\newacronym{d3}{D$^3$}{Deadline-Driven Delivery}
\newacronym{dac}{DAC}{Digital to Analog Converter}
\newacronym{dag}{DAG}{Directed Acyclic Graph}
\newacronym{darpa}{DARPA}{Defense Advanced Research Projects Agency}
\newacronym{das}{DAS}{Distributed Antenna System}
\newacronym{dash}{DASH}{Dynamic Adaptive Streaming over HTTP}
\newacronym{dc}{DC}{Dual Connectivity}
\newacronym{dccp}{DCCP}{Datagram Congestion Control Protocol}
\newacronym{dce}{DCE}{Direct Code Execution}
\newacronym{dci}{DCI}{Downlink Control Information}
\newacronym{dcl}{DCL}{Dear Colleague Letter}
\newacronym{dctcp}{DCTCP}{Data Center TCP}
\newacronym{devops}{DevOps}{Development and Operations}
\newacronym{dl}{DL}{Deep Learning}

\newacronym{dmr}{DMR}{Deadline Miss Ratio}
\newacronym{dmrs}{DMRS}{DeModulation Reference Signal}
\newacronym{drb}{DRB}{Data Radio Bearer}
\newacronym{drlcc}{DRL-CC}{Deep Reinforcement Learning Congestion Control}
\newacronym{drs}{DRS}{Discovery Reference Signal}
\newacronym{dt}{DT}{Digital Twin}
\newacronym{dtn}{DTN}{Digital Twin Network}
\newacronym{dtmn}{DTMN}{Digital Twin for Mobile Network}
\newacronym{dtwn}{DTWN}{Digital Twin Wireless Network}
\newacronym{du}{DU}{Distributed Unit}
\newacronym{e2e}{E2E}{end-to-end}
\newacronym{e2ap}{E2AP}{E2 Application Protocol}
\newacronym{e2sm}{E2SM}{E2 Service Model}
\newacronym{ecaas}{ECaaS}{Edge-Cloud-as-a-Service}
\newacronym{ecn}{ECN}{Explicit Congestion Notification}
\newacronym{edf}{EDF}{Earliest Deadline First}
\newacronym{em}{EM}{Electro-Magnetic}
\newacronym{embb}{eMBB}{Enhanced Mobile Broadband}
\newacronym{empower}{EMPOWER}{EMpowering transatlantic PlatfOrms for advanced WirEless Research}
\newacronym{enb}{eNB}{evolved Node Base}
\newacronym{endc}{EN-DC}{E-UTRAN-\gls{nr} \gls{dc}}
\newacronym{epc}{EPC}{Evolved Packet Core}
\newacronym{eps}{EPS}{Evolved Packet System}
\newacronym{es}{ES}{Edge Server}
\newacronym{etsi}{ETSI}{European Telecommunications Standards Institute}
\newacronym[firstplural=Estimated Times of Arrival (ETAs)]{eta}{ETA}{Estimated Time of Arrival}
\newacronym{eutran}{E-UTRAN}{Evolved Universal Terrestrial Access Network}
\newacronym{faas}{FaaS}{Function-as-a-Service}
\newacronym{fapi}{FAPI}{Functional Application Platform Interface}
\newacronym{fcc}{FCC}{Federal Communications Commission}
\newacronym{fdd}{FDD}{Frequency Division Duplexing}
\newacronym{fdm}{FDM}{Frequency Division Multiplexing}
\newacronym{fdma}{FDMA}{Frequency Division Multiple Access}
\newacronym{fed4fire}{FED4FIRE+}{Federation 4 Future Internet Research and Experimentation Plus}
\newacronym{fir}{FIR}{Finite Impulse Response}
\newacronym{fit}{FIT}{Future \acrlong{iot}}
\newacronym{fl}{FL}{Federated Learning}
\newacronym{fpga}{FPGA}{Field Programmable Gate Array}
\newacronym{fr2}{FR2}{Frequency Range 2}
\newacronym{fs}{FS}{Fast Switching}
\newacronym{fscc}{FSCC}{Flow Sharing Congestion Control}
\newacronym{ftp}{FTP}{File Transfer Protocol}
\newacronym{fw}{FW}{Flow Window}
\newacronym{ga128}{Ga}{Golay Sequence type A}
\newacronym{ge}{GE}{Gaussian Elimination}
\newacronym{glfsr}{GLFSR}{Galois Linear Feedback Shift Register}
\newacronym{gnb}{gNB}{Next Generation Node Base}
\newacronym{gold}{Gold}{Gold}
\newacronym{gop}{GOP}{Group of Pictures}
\newacronym{gpr}{GPR}{Gaussian Process Regressor}
\newacronym{gpu}{GPU}{Graphics Processing Unit}
\newacronym{gtp}{GTP}{GPRS Tunneling Protocol}
\newacronym{gtpc}{GTP-C}{GPRS Tunnelling Protocol Control Plane}
\newacronym{gtpu}{GTP-U}{GPRS Tunnelling Protocol User Plane}
\newacronym{gtpv2c}{GTPv2-C}{\gls{gtp} v2 - Control}
\newacronym{gw}{GW}{Gateway}
\newacronym{harq}{HARQ}{Hybrid Automatic Repeat reQuest}
\newacronym{hetnet}{HetNet}{Heterogeneous Network}
\newacronym{hh}{HH}{Hard Handover}
\newacronym{hol}{HOL}{Head-of-Line}
\newacronym{hqf}{HQF}{Highest-quality-first}
\newacronym{hss}{HSS}{Home Subscription Server}
\newacronym{http}{HTTP}{HyperText Transfer Protocol}
\newacronym{ia}{IA}{Initial Access}
\newacronym{iab}{IAB}{Integrated Access and Backhaul}
\newacronym{ic}{IC}{Incident Command}
\newacronym{ietf}{IETF}{Internet Engineering Task Force}
\newacronym{ifw}{IFW}{Interference Free Window}
\newacronym{imsi}{IMSI}{International Mobile Subscriber Identity}
\newacronym{imt}{IMT}{International Mobile Telecommunication}
\newacronym{iot}{IoT}{Internet of Things}
\newacronym{ip}{IP}{Internet Protocol}
\newacronym{iq}{IQ}{In-phase and Quadrature}
\newacronym{isi}{ISI}{Inter-Symbol Interference}
\newacronym{itu}{ITU}{International Telecommunication Union}
\newacronym{kpi}{KPI}{Key Performance Indicator}
\newacronym{kpm}{KPM}{Key Performance Measurement}
\newacronym{kvm}{KVM}{Kernel-based Virtual Machine}
\newacronym{lfsr}{LFSR}{Linear Feedback Shift Register}
\newacronym{los}{LOS}{Line-of-Sight}
\newacronym{ls}{LS}{Loosely Synchronised}
\newacronym{lsm}{LSM}{Link-to-System Mapping}
\newacronym{lstm}{LSTM}{Long Short Term Memory}
\newacronym{lte}{LTE}{Long Term Evolution}
\newacronym{lxc}{LXC}{Linux Container}
\newacronym{m2m}{M2M}{Machine to Machine}
\newacronym{mac}{MAC}{Medium Access Control}
\newacronym{mai}{MAI}{Multiple Access Interference}
\newacronym{manet}{MANET}{Mobile Ad Hoc Network}
\newacronym{mano}{MANO}{Management and Orchestration}
\newacronym{mc}{MC}{Multi-Connectivity}
\newacronym{mcc}{MCC}{Mobile Cloud Computing}
\newacronym{mchem}{MCHEM}{Massive Channel Emulator}
\newacronym{mcs}{MCS}{Modulation and Coding Scheme}
\newacronym{mec}{MEC}{Multi-access Edge Computing}
\newacronym{mec2}{MEC}{Mobile Edge Cloud}
\newacronym{mec3}{MEC}{Mobile Edge Computing}
\newacronym{mfc}{MFC}{Mobile Fog Computing}
\newacronym{mi}{MI}{Mutual Information}
\newacronym{mib}{MIB}{Master Information Block}
\newacronym{miesm}{MIESM}{Mutual Information Based Effective SINR}
\newacronym{mimo}{MIMO}{Multiple Input Multiple Output}
\newacronym{mgen}{MGEN}{Multi-Generator}
\newacronym{ml}{ML}{Machine Learning}
\newacronym{mlr}{MLR}{Maximum-local-rate}
\newacronym[plural=\gls{mme}s,firstplural=Mobility Management Entities (MMEs)]{mme}{MME}{Mobility Management Entity}
\newacronym{mmtc}{mMTC}{Massive Machine-Type Communications}
\newacronym{mmwave}{mmWave}{millimeter wave}
\newacronym{mpdccp}{MP-DCCP}{Multipath Datagram Congestion Control Protocol}
\newacronym{mptcp}{MPTCP}{Multipath TCP}
\newacronym{mr}{MR}{Maximum Rate}
\newacronym{mrdc}{MR-DC}{Multi \gls{rat} \gls{dc}}
\newacronym{mse}{MSE}{Mean Square Error}
\newacronym{mss}{MSS}{Maximum Segment Size}
\newacronym{mt}{MT}{Mobile Termination}
\newacronym{mtd}{MTD}{Machine-Type Device}
\newacronym{mtu}{MTU}{Maximum Transmission Unit}
\newacronym{mumimo}{MU-MIMO}{Multi-user \gls{mimo}}
\newacronym{mvno}{MVNO}{Mobile Virtual Network Operator}
\newacronym{nalu}{NALU}{Network Abstraction Layer Unit}
\newacronym{nas}{NAS}{Network Attached Storage}
\newacronym{nbiot}{NB-IoT}{Narrow Band IoT}
\newacronym{nf}{NF}{Network Function}
\newacronym{nfv}{NFV}{Network Function Virtualization}
\newacronym{nfvi}{NFVI}{Network Function Virtualization Infrastructure}
\newacronym{nic}{NIC}{Network Interface Card}
\newacronym{nlos}{NLOS}{Non-Line-of-Sight}
\newacronym{now}{NOW}{Non Overlapping Window}
\newacronym{nrdz}{NRDZ}{National Radio Dynamic Zone}
\newacronym{nsf}{NSF}{National Science Foundation}
\newacronym{nsm}{NSM}{Network Service Mesh}
\newacronym[type=hidden]{nr}{NR}{New Radio}
\newacronym{nrf}{NRF}{Network Repository Function}
\newacronym{nsa}{NSA}{Non Stand Alone}
\newacronym{nse}{NSE}{Network Slicing Engine}
\newacronym{nssf}{NSSF}{Network Slice Selection Function}
\newacronym{nssai}{NSSAI}{Network Slice Selection Assistance Information}
\newacronym{snssai}{S-NSSAI}{Single Network Slice Selection Assistance Information}
\newacronym{ntp}{NTP}{Network Time Protocol}
\newacronym{o2i}{O2I}{Outdoor to Indoor}
\newacronym{oai}{OAI}{OpenAirInterface}
\newacronym{oaicn}{OAI-CN}{\gls{oai} \acrlong{cn}}
\newacronym{oairan}{OAI-RAN}{\acrlong{oai} \acrlong{ran}}
\newacronym{oam}{OAM}{Operations, Administration and Maintenance}
\newacronym[plural=\gls{obu}s,firstplural=Onboard Units (OBUs)]{obu}{OBU}{Onboard Unit}
\newacronym{ofdm}{OFDM}{Orthogonal Frequency Division Multiplexing}
\newacronym{olia}{OLIA}{Opportunistic Linked Increase Algorithm}
\newacronym{omec}{OMEC}{Open Mobile Evolved Core}
\newacronym{onap}{ONAP}{Open Network Automation Platform}
\newacronym{onf}{ONF}{Open Networking Foundation}
\newacronym{onos}{ONOS}{Open Networking Operating System}
\newacronym{oom}{OOM}{\gls{onap} Operations Manager}
\newacronym{opnfv}{OPNFV}{Open Platform for \gls{nfv}}
\newacronym{orbit}{ORBIT}{Open-Access Research Testbed for Next-Generation Wireless Networks}
\newacronym{os}{OS}{Operating System}
\newacronym{osc}{OSC}{O-RAN Software Community}
\newacronym{osm}{OSM}{Open Street Map}
\newacronym{oss}{OSS}{Operations Support System}
\newacronym{pa}{PA}{Position-aware}
\newacronym{pase}{PASE}{Prioritization, Arbitration, and Self-adjusting Endpoints}
\newacronym{pawr}{PAWR}{Platforms for Advanced Wireless Research}
\newacronym{pbch}{PBCH}{Physical Broadcast Channel}
\newacronym{pcef}{PCEF}{Policy and Charging Enforcement Function}
\newacronym{pcfich}{PCFICH}{Physical Control Format Indicator Channel}
\newacronym{pcrf}{PCRF}{Policy and Charging Rules Function}
\newacronym{pdcch}{PDCCH}{Physical Downlink Control Channel}
\newacronym{pdcp}{PDCP}{Packet Data Convergence Protocol}
\newacronym{pdsch}{PDSCH}{Physical Downlink Shared Channel}
\newacronym{pdu}{PDU}{Packet Data Unit}
\newacronym{pdp}{PDP}{Power Delay Profile}
\newacronym{pf}{PF}{Proportional Fair}
\newacronym{pgw}{PGW}{Packet Gateway}
\newacronym{phich}{PHICH}{Physical Hybrid ARQ Indicator Channel}
\newacronym{phy}{PHY}{Physical}
\newacronym{pl}{PL}{Path Loss}
\newacronym{plmn}{PLMN}{Public Land Mobile Network}
\newacronym{pmch}{PMCH}{Physical Multicast Channel}
\newacronym{pmi}{PMI}{Precoding Matrix Indicators}
\newacronym{powder}{POWDER}{Platform for Open Wireless Data-driven Experimental Research}
\newacronym{ppo}{PPO}{Proximal Policy Optimization}
\newacronym{ppp}{PPP}{Poisson Point Process}
\newacronym{prach}{PRACH}{Physical Random Access Channel}
\newacronym{prb}{PRB}{Physical Resource Block}
\newacronym{psnr}{PSNR}{Peak Signal to Noise Ratio}
\newacronym{pss}{PSS}{Primary Synchronization Signal}
\newacronym{pucch}{PUCCH}{Physical Uplink Control Channel}
\newacronym{pusch}{PUSCH}{Physical Uplink Shared Channel}
\newacronym{qam}{QAM}{Quadrature Amplitude Modulation}
\newacronym{qci}{QCI}{\gls{qos} Class Identifier}
\newacronym{qoe}{QoE}{Quality of Experience}
\newacronym{qos}{QoS}{Quality of Service}
\newacronym{qtgui}{QT-GUI}{QT Graphical User Interface}
\newacronym{quic}{QUIC}{Quick UDP Internet Connections}
\newacronym{rach}{RACH}{Random Access Channel}
\newacronym{ran}{RAN}{Radio Access Network}
\newacronym[firstplural=Radio Access Technologies (RATs)]{rat}{RAT}{Radio Access Technology}
\newacronym{rb}{RB}{Resource Block}
\newacronym{rcn}{RCN}{Research Coordination Network}
\newacronym{rec}{REC}{Radio Edge Cloud}
\newacronym{red}{RED}{Random Early Detection}
\newacronym{renew}{RENEW}{Reconfigurable Eco-system for Next-generation End-to-end Wireless}
\newacronym{rf}{RF}{Radio Frequency}
\newacronym{rfsim}{RFSim}{RF Simulation}
\newacronym{rfc}{RFC}{Request for Comments}
\newacronym{rfr}{RFR}{Random Forest Regressor}
\newacronym{ric}{RIC}{RAN Intelligent Controller}
\newacronym{rlc}{RLC}{Radio Link Control}
\newacronym{rlf}{RLF}{Radio Link Failure}
\newacronym{rlnc}{RLNC}{Random Linear Network Coding}
\newacronym{rmse}{RMSE}{Root Mean Squared Error}
\newacronym{rnis}{RNIS}{Radio Network Information Service}
\newacronym{rnti}{RNTI}{Radio Network Temporary Identifier}
\newacronym{rr}{RR}{Round Robin}
\newacronym{rrc}{RRC}{Radio Resource Control}
\newacronym{rrm}{RRM}{Radio Resource Management}
\newacronym{rru}{RRU}{Remote Radio Unit}
\newacronym{rs}{RS}{Remote Server}
\newacronym{rsc}{RSC}{RAN Slicing Control}
\newacronym{rsrp}{RSRP}{Reference Signal Received Power}
\newacronym{rsrq}{RSRQ}{Reference Signal Received Quality}
\newacronym{rss}{RSS}{Received Signal Strength}
\newacronym{rssi}{RSSI}{Received Signal Strength Indicator}
\newacronym{rsu}{RSU}{Road-Side Unit}
\newacronym{rtt}{RTT}{Round Trip Time}
\newacronym{ru}{RU}{Radio Unit}
\newacronym{rw}{RW}{Receive Window}
\newacronym{rx}{RX}{Receiver}
\newacronym{s1ap}{S1AP}{S1 Application Protocol}
\newacronym{sa}{SA}{standalone}
\newacronym{sack}{SACK}{Selective Acknowledgment}
\newacronym{sap}{SAP}{Service Access Point}
\newacronym{sc2}{SC2}{Spectrum Collaboration Challenge}
\newacronym{scef}{SCEF}{Service Capability Exposure Function}
\newacronym{sch}{SCH}{Secondary Cell Handover}
\newacronym{scoot}{SCOOT}{Split Cycle Offset Optimization Technique}
\newacronym{sctp}{SCTP}{Stream Control Transmission Protocol}
\newacronym{sdap}{SDAP}{Service Data Adaptation Protocol}
\newacronym{sd}{SD}{Slice Differentiator}
\newacronym{sdk}{SDK}{Software Development Kit}
\newacronym{sdm}{SDM}{Space Division Multiplexing}
\newacronym{sdma}{SDMA}{Spatial Division Multiple Access}
\newacronym{sdn}{SDN}{Software-defined Networking}
\newacronym{sdr}{SDR}{Software-defined Radio}
\newacronym{seba}{SEBA}{SDN-Enabled Broadband Access}
\newacronym{sgsn}{SGSN}{Serving GPRS Support Node}
\newacronym{sgw}{SGW}{Service Gateway}
\newacronym{si}{SI}{Study Item}
\newacronym{sib}{SIB}{Secondary Information Block}
\newacronym{sinr}{SINR}{Signal to Interference plus Noise Ratio}
\newacronym{sip}{SIP}{Session Initiation Protocol}
\newacronym{siso}{SISO}{Single Input, Single Output}
\newacronym{sla}{SLA}{Service Level Agreement}
\newacronym{sm}{SM}{Service Model}
\newacronym{smf}{SMF}{Session Management Function}
\newacronym{smo}{SMO}{Service Management and Orchestration}
\newacronym{sms}{SMS}{Short Message Service}
\newacronym{smsgmsc}{SMS-GMSC}{\gls{sms}-Gateway}
\newacronym{snr}{SNR}{Signal-to-Noise-Ratio}
\newacronym{son}{SON}{Self-Organizing Network}
\newacronym{sptcp}{SPTCP}{Single Path TCP}
\newacronym{srb}{SRB}{Signalling Radio Bearer}
\newacronym{srn}{SRN}{Standard Radio Node}
\newacronym{srs}{SRS}{Sounding Reference Signal}
\newacronym{ss}{SS}{Synchronization Signal}
\newacronym{sss}{SSS}{Secondary Synchronization Signal}
\newacronym{sst}{SST}{Slice/Service Type}
\newacronym{st}{ST}{Spanning Tree}
\newacronym{svc}{SVC}{Scalable Video Coding}
\newacronym{tb}{TB}{Transport Block}
\newacronym{tbs}{TBS}{Transport Block Size}
\newacronym{tcp}{TCP}{Transmission Control Protocol}
\newacronym{tdd}{TDD}{Time Division Duplexing}
\newacronym{tdm}{TDM}{Time Division Multiplexing}
\newacronym{tdma}{TDMA}{Time Division Multiple Access}
\newacronym{tfl}{TfL}{Transport for London}
\newacronym{tfrc}{TFRC}{TCP-Friendly Rate Control}
\newacronym{tft}{TFT}{Traffic Flow Template}
\newacronym{tgen}{TGEN}{Traffic Generator}
\newacronym{tip}{TIP}{Telecom Infra Project}
\newacronym{tm}{TM}{Transparent Mode}
\newacronym{to}{TO}{Telecom Operator}
\newacronym{toa}{ToA}{Time of Arrival}
\newacronym{tr}{TR}{Technical Report}
\newacronym{trp}{TRP}{Transmitter Receiver Pair}
\newacronym{ts}{TS}{Technical Specification}
\newacronym{tti}{TTI}{Transmission Time Interval}
\newacronym{ttt}{TTT}{Time-to-Trigger}
\newacronym{tx}{TX}{Transmitter}
\newacronym{uas}{UAS}{Unmanned Aerial System}
\newacronym{uav}{UAV}{Unmanned Aerial Vehicle}
\newacronym{udm}{UDM}{Unified Data Management}
\newacronym{udp}{UDP}{User Datagram Protocol}
\newacronym{udr}{UDR}{Unified Data Repository}
\newacronym{ue}{UE}{User Equipment}
\newacronym{uhd}{UHD}{\gls{usrp} Hardware Driver}
\newacronym{ul}{UL}{Uplink}
\newacronym{um}{UM}{Unacknowledged Mode}
\newacronym{uml}{UML}{Unified Modeling Language}
\newacronym{upa}{UPA}{Uniform Planar Array}
\newacronym{upf}{UPF}{User Plane Function}
\newacronym{urllc}{URLLC}{Ultra Reliable and Low Latency Communication}
\newacronym{usa}{U.S.}{United States}
\newacronym{usim}{USIM}{Universal Subscriber Identity Module}
\newacronym{usrp}{USRP}{Universal Software Radio Peripheral}
\newacronym{utc}{UTC}{Urban Traffic Control}
\newacronym{vim}{VIM}{Virtualization Infrastructure Manager}
\newacronym{vm}{VM}{Virtual Machine}
\newacronym{vnf}{VNF}{Virtual Network Function}
\newacronym{volte}{VoLTE}{Voice over \gls{lte}}
\newacronym{voltha}{VOLTHA}{Virtual OLT HArdware Abstraction}
\newacronym{vr}{VR}{Virtual Reality}
\newacronym{vran}{vRAN}{Virtualized \gls{ran}}
\newacronym{vss}{VSS}{Video Streaming Server}
\newacronym{wbf}{WBF}{Wired Bias Function}
\newacronym{wf}{WF}{Wired-first}
\newacronym{wi}{WI}{Wireless InSite}
\newacronym{wlan}{WLAN}{Wireless Local Area Network}
\newacronym{pnf}{PNF}{Physical Network Function}
\newacronym{drl}{DRL}{Deep Reinforcement Learning}
\newacronym{mtc}{MTC}{Machine-type Communications}
\newacronym{v2x}{V2X}{Vehicle-to-everything}
\newacronym{cast}{CaST}{Channel emulation scenario generator and Sounder Toolchain}
\newacronym{gui}{GUI}{Graphical User Interface}
\newacronym{ups}{UPS}{Uninterruptible Power Supply}
\newacronym{ota}{OTA}{Over-the-Air}
\newacronym{hitl}{HITL}{hardware-in-the-loop}
\newacronym{soc}{SoC}{System-on-Chip}
\newacronym{cnn}{CNN}{Convolutional Neural Network}
\newacronym{nlp}{NLP}{natural language processing}
\newacronym{nn}{NN}{Neural Network}
\newacronym{onnx}{ONNX}{Open Neural Network Exchange}
\newacronym{dft}{DFT}{discrete Fourier transform}
\newacronym{flops}{FLOPs}{floating point operations}
\newacronym{nmse}{NMSE}{normalized mean-square error}
\newacronym{cdl}{CDL}{clustered delay line}